\documentclass[fleqn,twoside]{article}
\usepackage{espcrc2}

\usepackage{epsfig}

\newcommand{\nuance}{\texttt{nuance}}
\newcommand{\Nuance}{\texttt{Nuance}}
\newcommand{\Oxygen}{\ensuremath{{}^{16}\mathrm{O}}}

\title{The \nuance\ Neutrino Physics Simulation, and the Future}

\author{D. Casper\address{Department of Physics and Astronomy \\
University of California, Irvine \\
Irvine, CA 92697, USA \\
\texttt{dcasper@uci.edu}}}

\begin{document}

\begin{abstract}
This article briefly describes the \nuance\ neutrino simulation software and outlines
the program of the working group on neutrino event generators which met for the first time at
the NUINT'01 meeting.

\end{abstract}

\maketitle

\section{HISTORY AND OVERVIEW}

The unrequited search for proton decay with large underground detectors,
begun almost 20 years ago, focused attention on describing atmospheric neutrino
reactions in detail.\cite{IMB1,HainesThesis,KamMc} Unlike typical accelerator beams of neutrinos, the spectrum of atmospheric
neutrinos covers many decades of baseline and energy.  From the lowest-energy contained
interactions to the highest-energy upward-going muons, neutrino energies
in a broad range ($\sim$100 MeV to $>$1 TeV) are sampled with roughly comparable rates.
These features, which would eventually help reveal neutrino mass and oscillation, demand
a model of exclusive reactions valid for essentially any interacting
neutrino energy.\cite{DwcThesis,IMB3}

With the advent of Super-Kamiokande, the statistical precision of atmospheric neutrino measurements
increased dramatically and systematic uncertainties (fluxes, cross-sections and calibration) became
increasingly important.  The same situation is likely to arise in future long-baseline
experiments with high-luminosity neutrino beams.

The \nuance\ software package described here is the author's work-in-progress on the problem of modeling neutrino
interactions.\footnote{This write-up reflects version 2.000 of the code.} Although originally used for atmospheric
neutrino interactions, the program was designed to be useful in as many other
applications as possible.

\section{NEUTRINO PHYSICS MODEL}

Space does not permit a full elaboration of the cross-section calculations for
each reaction channel or graphical comparisons with experimental data,
so a schematic outline with references must suffice.

\subsection{Generalities}

\Nuance\ adopts a ``divide and conquer'' strategy.  Models (with
varying degrees of sophistication) can be found in the literature for each general class of reaction. By summing the cross-sections
and rates of all exclusive channels, and then adding (inclusive) deep-inelastic scattering inside
appropriate kinematic limits, the total cross-sections and event rates are obtained.

The units used by the program are MeV, grams (for densities), centimeters and seconds.  The user does not directly
interact with cross-sections, but these are stored internally in units of $10^{-36} \ \hbox{cm}^2$, or picobarns.

Simulated elementary particles are indexed using \texttt{PYTHIA}\cite{Pythia6}/Particle Data Group\cite{PDG} codes. Application-defined codes
are used for nuclei in the few cases they are needed.  Particle properties and physical constants are generally taken from
\texttt{PYTHIA} internal stuctures by default, although any parameter can be overridden at run-time.
Where more recent values are available, they are taken from the Review of Particle Properties.  The \texttt{PYTHIA} particle
stack is used to build the event in memory and process some decays.

The Fermi gas model is used to simulate the effect of a bound nucleon target (Fermi motion and Pauli blocking). The implementation of this
model is slightly different from process to process, however in general bound nucleons are given a uniform initial momentum {\it density} up to
a (user-specified) maximum value, and a negative binding energy.  While different shells (with different momenta and binding energy)
are supported, in practice the values $p_f \le 225 \ \hbox{MeV/c}$ and $E_b = -27 \ \hbox{MeV}$ (interpolated from a fit
to electron-scattering data\cite{Moniz:1971mt}) are used for all 16 nucleons in the canonical case of \Oxygen.
A final state nucleon must exceed a threshold momentum value to exit the nucleus and allow the reaction to
occur.  In the absence of final-state interaction, the (observable) outgoing nucleon momentum must therefore be greater than 225~MeV/c. Discussions at the
Workshop made it abundantly clear that this hard cut-off is unphysical, however a more satisfactory presciption remains elusive.  In most cases,
nucleons below 225 MeV/c will not be detectable, but prudence is advised in interpreting low-energy recoil nucleons.

Lepton masses are never neglected, and the appropriate vector-boson mass is always included in the propagator.

\subsection{Electron Scattering}

The only exactly-calculable (at first-order) neutrino cross-section is unfortunately also the least important for neutrino energies
above a few tens of MeV. For completeness, all purely leptonic neutrino and anti-neutrino reactions are treated, including elastic scattering
and inverse muon-decay \mbox{($\overline{\nu}_{e} e^{-} \rightarrow \mu^{-} \overline{\nu}_{\mu}$)}. The tree-level cross-section formul\ae\ are given in many particle physics textbooks.\cite{Greiner}

\subsection{Quasi-elastic Scattering}

In \nuance\ quasi-elastic scattering comprises both charged- and neutral-current two-body neutrino reactions with nucleons.
The relativistic Fermi gas model of Smith and Moniz provides the general framework for all such processes.\cite{Smith:xh}
To ensure consistency between bound and free targets, free nucleon cross-sections are also calculated using the Smith-Moniz formalism, setting the binding energy equal to zero and expressing the initial nucleon momentum distribution as a delta function at zero.

Identical form factors are used for both free and bound nucleons.  The usual dipole parameterization of vector and
axial-vector form factors is adopted, with default values $m_V = 0.840\ \mbox{GeV}/c^2$ and $m_A = 1.00\ \mbox{GeV}/c^2$, respectively. For the induced pseudoscalar form factor, a parameterization calculated from lattice QCD\cite{Liu:1994dr} (which agrees extremely well with low-$q^2$ pion electroproduction data) is used by default, although the simpler pion-pole expression from \cite{Smith:xh} can be selected instead. Second-class currents are assumed to vanish, as required by the $V-A$ electroweak theory and fundamental symmetries.

The total and differential charged-current cross-sections of the Smith-Moniz model agree well with more sophisticated theoretical calculations for neutrino energies above 50-100~MeV where continuum excitation of \Oxygen\ is dominant.

Neutral-current two-body reactions with nucleons are usually invisible, but have been incorporated into the framework of the Smith-Moniz model and included for completeness. Neutral-current nucleon form factors are specified by the electroweak theory and can be related to those for
charged-current reactions.\cite{Ahrens:xe}

Similarly, the Smith-Moniz model has been extended to include charged-current, Cabibbo-suppressed hyperon production, following the treatment of Pais\cite{Pais} to account for the inelasticity of such reactions and the $|\Delta I| = \frac{1}{2}$ rule.

\subsection{Resonant Processes}

For neutrino energies around 1~GeV and above, baryon resonances may be excited and subsequently decay into a nucleon and one or more mesons.  The most prominent resonance is the $\Delta\textrm{(1232)}$, but $N\mathrm{(1440)}$ and a number of others can also contribute.  Rein and Sehgal\cite{ReinSehgal} have shown that these resonance-mediated channels can be described using harmonic oscillator quark wavefunctions, symmetrized under SU(6) (extending the approach of Feynman, Kislinger and Ravndal\cite{Feynman:wr} to account for higher-mass resonant states and the isospin structure of the weak interaction). For hadronic masses above the $\Delta\textrm{(1232)}$, the data appear to require inclusion of interference between resonances with identical isospin. \Nuance\ adopts the Rein-Sehgal model, modified to account for improved knowledge of the mass spectrum since publication of their original paper, and incorporates non-strange resonances up to 2~GeV. Nucleon form factors for resonance production are assumed identical to those for quasi-elastic scattering, however Rein and Seghal neglect the pseudoscalar form factor of the nucleon, which (in principle) could be important for $\nu_{\tau}$ charged-current reactions.\footnote{In practice, deep-inelastic scattering is the dominant $\nu_{\tau}$ charged-current reaction channel for most kinematically interesting neutrino energies.}

Rein and Sehgal consider only resonance decays into $N \pi$ final states, but additional channels are possible. For completeness, the program attempts to model these by simply adding the additional decay modes so that the total branching fraction for each resonance sums to 100\%.  The procedure assumes the decay matrix element for other modes is identical to the $N \pi$ case, which is probably wrong, but the contribution of these more exotic reactions to the total cross-section is extremely small.  Unfortunately, due to the complexity of the Rein-Sehgal calculation, the large number of independent integration variables, and the increased number of channels, the bulk of the program's computing time is spent on these relatively unimportant reactions.

For reactions on bound nucleons, the initial state is given a uniform Fermi momentum density and a negative binding energy. Reactions in which resonance decay would result in a nucleon below the Fermi sea are Pauli-blocked and do not occur.  Kim et al.\cite{Kim:1996az}, Singh and collaborators\cite{Alvarez-Ruso:1999dy} and Marteau et al.\cite{Marteau:ur} have pointed out
that the nuclear medium can modify the widths of resonances and allow the final-state reaction $N^* N \rightarrow N N$. In-medium effects on the widths of resonances are not considered by the program, but the ``pion-less $\Delta$ decay" reaction (a misnomer, since it seems relevant to $I=\frac{1}{2}$ channels as well) can reduce the number of pions produced by 10-50\% (the numbers suggested in the literature vary considerably). It is difficult to disentangle $N^* N \rightarrow N N$ from similar processes in which $N^* \rightarrow N \pi$ decay occurs but the pion is subsequently absorbed, and it is not clear to what extent the former reaction is double-counted by final-state interactions.  Nevertheless, Super-Kamiokande and K2K data appear to favor some suppression of resonant pion production.  Absent a theoretically well-motivated and consistent description of $N^* N \rightarrow N N$ in the Fermi gas model, by default \nuance\ adopts an {\it ad hoc} 20\% suppression of pion production for $I_3 = \pm \frac{1}{2}$ reactions and a 10\% suppression for $I_3 = \pm \frac{3}{2}$.\footnote{Na\"{\i}vely, a $\Delta^{++}$ or $\Delta^-$ can only participate in the $N^* N \rightarrow N N$ reaction with half as many nucleons as $\Delta^+$ and $\Delta^0$.} Hopefully interaction between the nuclear- and neutrino-physics communities fostered by the NUINT workshop will help clarify this important effect.

\subsection{Coherent and Diffractive Reactions}

In coherent reactions, neutrinos scatter from an entire nucleus rather than its individual constituents, with negligible energy transfer to the target. Coherent reactions typically produce a forward-going lepton and meson; both charged- and neutral-current reactions are possible, leading to charged or neutral meson production, respectively. As a purely axial-vector process, the cross-sections for neutrinos and anti-neutrinos are equal, and the neutral-current cross-section is half as large as for charged currents.  

\Nuance\ implements Rein and Sehgal's calculation\cite{Rein:1982pf} of the cross-section for coherent pion-production in terms of the $\pi A$ forward scattering cross-section. Rein and Sehgal's cross-section agrees with the handful of measurements available, however the lowest-energy data available is at $E_{\nu} =  \ \mathrm{2 \, GeV}$, with an Al target.  Using a far more detailed model of the nuclear physics, Kelkar et al.\cite{Kelkar:1996iv}
have stressed the importance of nuclear medium effects on the $\Delta$ which mediates these reactions, and predict a dramatic suppression of the coherent cross-section for neutrino energies around 1~GeV.  Even with the larger cross-sections predicted by Rein and Sehgal, coherent pion production represents a small fraction of the incoherent, resonance-mediated pion-production cross-section for energies around 1~GeV, so the impact of this effect on the total pion production rate should be 10\% or less. Data from the near detectors of K2K, and eventually MiniBooNE, should allow the coherent contribution to be separated from the resonant channels based on the observed single-pion angular distribution.

Diffractive reactions in \nuance\ are analogous to coherent, except free protons rather than \Oxygen\ nuclei are the targets; the dynamics of the two reactions are identical.  Rein's calculation\cite{Rein:cd} of the diffractive pion-production cross-section, using a model similar to the coherent case, is implemented in \nuance\ by the same routines which handle coherent reactions.  The smaller size of the target reduces the cross-section  correspondingly, by approximately a factor $16^{\frac{2}{3}}$. Although the nuclear-medium effects cited in \cite{Kelkar:1996iv} do not apply to free nucleon targets, the cross-section for diffractive single-pion production is dwarfed by incoherent resonant channels and coherent reactions with \Oxygen, hence it is included mainly for completeness.

Coherent and diffractive production of vector mesons has been measured by several high-energy experiments, at the few per-mille level compared to the total charged-current rate. Cross-sections for these reactions, which are presently ignored in \nuance, have been estimated based on the CVC hypothesis and vector meson dominance and appear to describe the data reasonably well.\cite{vmd}  For completeness, these channels will be added in a future version of the program.

\subsection{Deep-inelastic Scattering}

Deep-inelastic scattering is unique in that it is modeled as an inclusive, rather than exclusive, reaction between a neutrino and the parton consitituents of the nucleon. Elementary formul\ae\ are derived in most particle physics texts, however the usual expressions involve a number of undesirable kinematic approximations, including neglect of lepton (and target, in some cases) masses and the induced pseudoscalar form factor.  The unpublished calculation of Roe\cite{Roe} avoids these limitations, and forms the kinematic basis for the treatment of this reaction in \nuance.  Additional complications arise at low-energies, where the threshold for production of real heavy quarks (c,b) is important.  The usual approach is to impose ``slow rescaling"; here the calculation in \nuance\ follows Leader and Predazzi.\cite{Leader:hm}

\Nuance\ interfaces to the \texttt{PDFLIB} package,\cite{Plothow-Besch:1992qj} allowing one of dozens of nucleon structure function parameterizations to be selected at run-time. By default, the program uses the BEBC\cite{BEBC} option, chosen because it was measured with neutrinos and has a relatively low $q^2$ cut-off. Care is required in dealing with structure functions provided by \texttt{PDFLIB} since requests for structure functions with $|q^2| < |q^2_{min}|$ result in the values for $q^2 = q^2_{min}$ being returned. In this case, the behavior of the structure functions as $q^2 \rightarrow 0$ is extrapolated according to the vector meson dominance model of \texttt{PYTHIA}.\cite{Pythia6}

Shadowing, anti-shadowing, Fermi and ``EMC" effects on the structure functions are neglected. Form factors are built from the structure functions assuming the Callen-Gross relation ($F_2 = 2xF_1$) is exact and thereby also neglecting the longitudinal cross-section.

Since the deep-inelastic cross-section includes resonant (and perhaps quasi-elastic) processes with low hadronic mass (which are already included via exclusive channels) the kinematic limits of integration must be chosen to avoid double-counting. As extensively discussed at NUINT'01, in real-life there is no distinct cut-off between ``resonant" and deep-inelastic scattering, but rather a smooth transition from scattering off hadrons to scattering off quarks, as the vector boson probe begins to resolve the internal structure of the target. The present version of the \nuance\ attempts to square this circle by integrating the deep-inelastic cross-section over the limits $|q^2| > 1 \ (\mathrm{GeV/c})^2 \ \mathrm{.OR.} \ W > 2 \ \mathrm{GeV}$.\footnote{The additional constraint $W > m_N + m_{\pi}$ is also imposed, regardless of $q^2$.}  These values are chosen to make the differential cross-section roughly continuous across the artificial boundary at $W = 2 \ \hbox{GeV}$, the upper limit for resonant reactions in the program, although it is only partially successful.

Because the deep-inelastic scattering cross-section is inclusive, these channels are different from others in the program, where the final-state particles are uniquely determined by the reaction itself. For deep-inelastic scattering, the interacting quark is determined by a weighting scheme from the \texttt{LEPTO} program\cite{Ingelman:1996mq} and transformed into a different flavor (for charged-current reactions). Parton showers and fragmentation of the outgoing quark are treated by a version of \texttt{LEPTO} which has been extensively modified to ensure conservation of energy and momentum and has kinematic cuts adjusted so $|q^2_{min}| = 10^{-3} \ (\mathrm{GeV/c})^2$ and $W^2_{min} = 2 \ \mathrm{GeV}^2$.  \texttt{LEPTO} itself uses \texttt{PYTHIA} for some of its work.

Because low-mass and small-$q^2$ reactions are sometimes generated, the LUND string fragmentation model occasionally fails to produce a final state. In this case KNO scaling\cite{Koba:ng} is used to determine a final-state using measured multiplicity distributions and fragmentation functions.\cite{Schmitz} 

Deep-inelastic scattering is undoubtedly the regime where \nuance\ currently lags behind the state-of-the-art in other generators dedicated to high-energy neutrino physics. The NUINT meeting has already borne fruit by bringing the issue of quark/hadron duality to the forefront and eliciting at least one clear prescription for how to remedy these shortcomings.\cite{Bodek:2002vp}  This approach will be implemented in a future version of the program.

\subsection{Nuclear Processes}

The program uses a model of final state interaction in the nucleus originally developed for the IMB experiment.\cite{Manjeet}
At present the model is specific to \Oxygen, although it could be adapted to other nuclei.  Primary interactions are
assigned a starting position in the nucleus according to the measured density distribution for \Oxygen.\cite{Sick} Hadrons are then
tracked through the nucleus in 0.2~fm steps, treating the nucleus as an isoscalar sphere of nuclear matter with radially-dependent density and
Fermi momentum.  Single-nucleon cross-sections and local density are used to calculate the interaction probability during each step. Interactions
resulting in a nucleon with momentum below the Fermi sea are Pauli-blocked and ignored.

Measured cross-sections and angular distributions are used for $\pi-N$ and $N-N$ reactions.\cite{CERN-HERA} Angular
distributions for elastic reactions are calculated from a global phase-shift analysis of world data.\cite{Karlsruhe}
Inelastic reactions involving up to five particles are possible. For three-body $\pi-N$ reactions, a $\Delta I = \frac{1}{2}$ dominance model
is adopted;\cite{OllsenYodh} in other cases, a simple resonance model is assumed.
For reactions without available data, cross-sections are inferred from isospin symmetry.

Interactions of kaons with kinetic energy up to 1~GeV are also simulated. Cross-sections and angular distributions for all
two-body $K-N$ reactions are calculated using a partial-wave analysis.\cite{Kaons} Neutral kaons are treated as
50\% $K_S$ (which decay immediately) and 50\% $K_L$. $K_L$ cross-sections are calculated as an equal mixture of
$K^0 \mathrm{and}\ \overline{K}^0$, and any interaction has a 50\% chance of regenerating a $K_S$ which (again) immediately decays.

Hyperon interactions inside the nucleus are not simulated and $\rho$ mesons decay rapidly before any interaction is possible. Elastic
scattering and pion-production cross-sections\cite{rhoOmega} compete with decay for $\omega$ and $\eta$ mesons inside the nucleus.

The IMB cascade model has been tested by running the algorithm in a mode which simulates $\pi,p,K - \Oxygen$ scattering by
starting a hadron {\it outside} the nucleus and stepping it through.  The pion absorption cross-section is tuned to reproduce
measurements; for other channels, the results of the simulation reproduce scattering data for hadrons on \Oxygen\ and $^{12}\mathrm{C}$ extremely well for kinetic energies up to 2~GeV. For pions, the model also agrees with the less detailed (but more elegant) analytical approach of Adler, Nussinov and Paschos.\cite{Adler:qu}

One flaw in such superposition models was pointed out at NUINT'01, namely the failure to account for ``formation zones", which suppress the
final-state interactions of hadrons with energies greater than a few GeV.\cite{fzic}  In the present version of the program, cross-sections are assumed
to be constant for energies above 2~GeV.

After ejection of a nucleon recoiling from a neutrino interaction, de-excitation and/or break-up of the residual nucleus can result in
emission of one or two
few-MeV $\gamma$ along with possible evaporation of low-energy nucleons.\cite{Ejiri} While the de-excitation products are usually quite
low in energy, their signature in a water detector is sufficient to shift the average reconstructed mass of $\pi^0$ produced in neutral current reactions
by as much as 5~MeV.  The model of de-excitation currently used is specific to \Oxygen, and should be generalized.

Once all neutrino interaction products have escaped the nucleus or been absorbed, and the nucleus has de-excited, \nuance\ completes its simulation of
the event by calling \texttt{PYTHIA} to process decays of short-lived particles outside the nucleus. 

\subsection{Utilities}

\Nuance\ provides additional services for specialized applications, and uses others which are publicly available. The program
includes a package for modulating input fluxes under 3-component neutrino oscillations in matter with CP-violation\cite{Barger} (including
a realistic ``onion-skin" model of the Earth's density profile). In practice, one often prefers to generate unoscillated (charged- plus neutral-current)
and ``fully-oscillated" (charged-current only; $P_{osc}=1$, independent of energy) samples, and then reweight them according to different
hypotheses after the fact.  The program will also generate nucleon decays in a user-specified channel, using the same
bound nucleon and final-state interaction model applied to neutrino interactions.

Decays of $\tau$ leptons are handled by the \texttt{TAUOLA} package.\cite{Tauola}  The polarization of primary $\tau$ leptons produced by neutrinos
is calculated using the form factors for quasi-elastic and deep-inelastic reactions.\cite{Albright:1974ts}
In resonant and diffractive processes,
the appropriate form factors are unclear and/or difficult to calculate, so the tau is assumed to be completely polarized.

The program can also simulate entering neutrino-induced muons produced in the material outside a detector. For transport of energetic muons
to the detector surface, the \texttt{PROPMU} package\cite{Lipari:ut} is used.  Note that small-angle Coulomb scattering of muons
in flight is neglected for computational efficiency.

A stand-alone utility program (\texttt{nuplot}) is also provided to translate the calculated cross-sections and rates from the internal format
of a sequential access \texttt{ZEBRA/FZ} file created by the program into an Ntuple for inspection. Using this utility, energy-dependent cross-sections for any of the many exclusive reactions can be individually plotted.

\subsection{Validity and Limitations}

Unphysical final-state nucleon energies in the Fermi gas model, the \Oxygen-specific description of final-state interactions and a number of other problems and simplifications have been discussed above.

Because the program is written in \texttt{FORTRAN77} and uses \texttt{ZEBRA}\cite{CernlibZebra} for dynamical memory management, 
single precision calculations are performed in most cases (\texttt{PYTHIA} uses double precision calculations internally).
This is an unsatisfactory solution, since it requires considerable care
in coding to preserve numerical precision (such problems typically appear for neutrino energies approaching 1~TeV).
A future version of the program, using a language with native dynamic allocation (such as
\texttt{FORTRAN90} or \texttt{C++}), will eventually remove this handicap.

\section{INTERFACE}

The program was intended to be flexible and easy to use, although these requirements are frequently at odds, if not mutually exclusive. To allow wider access to the code, it has been posted online at \texttt{http://nuint.ps.uci.edu/nuance}. The code is developed and maintained under the Windows operating system, but is fully compatible with others. The \texttt{CMT} configuration management tool\cite{CMT} provides a platform-independent mechanism to build the program and its constituent libraries automatically. The main program itself consists of only eight lines of executable code, hence the libraries that do the actual work can be easily slaved to an alternative user-written program.

The program is steered using a \texttt{ZEBRA/TZ}\cite{CernlibZebra} (text) cards file, although frequently-changed options can also be specified on the command line.  Default physical parameters are stored in a second cards file named \verb+nuance_defaults.cards+. Values specified in the job-specific file provided by the user override the defaults. The user can also modify any \texttt{PYTHIA} parameter via a data card.

\subsection{Setup}

The user must describe the geometry and composition of the target before running the program. The description syntax is
reminiscent of \texttt{GEANT}, but somewhat simpler.  A target consists of a one or more volumes (sphere, cylinder or box) which can be nested, but presently must all be centered at the origin. Each volume contains a mixture of one or more materials, with user-specified density.  Materials are in turn composed of atoms, which are built from neutrons and protons in shells with specific Fermi momenta and binding energy.  Each atom is assumed to include a number of electrons equal to the number of protons.

\subsection{Fluxes}

The program can generate events for a mono-energetic, single-flavor beam of neutrinos, however for most applications the user must also provide a flux of neutrinos.  Drivers compatible with several atmospheric flux calculation formats are included.

An accelerator beam can be described via an \texttt{HBOOK} file containing one or more histograms (each associated with a particular neutrino flavor) in units of neutrinos per bin per cm$^2$ per unit luminosity. These histograms must have identical upper and lower limits, but the user may specify whether the limits are interpreted as MeV or GeV. Two scaling factors, one linear and one quadratic, allow the luminosity units and baseline of the beam (respectively) to be changed at run-time. The beam direction in local coordinates is also adjustable.  The user may apply multi-quadric smoothing on the input histogram to eliminate binning effects, or directly generate neutrinos from the raw values.

Finally, a simple model of thermal neutrino emission (for supernov\ae)\cite{Spergel:1987ch}
is built into program, allowing the user to specify a total luminosity and cooling time for one or more neutrino flavors. In this special case, the fluxes are time-dependent and interaction rates are recalculated as the simulation progresses and the source cools.

\subsection{Cross-sections and Rates}

Having described what is to simulated, the user must calculate neutrino cross-sections and interaction rates. The caculation
is time-consuming (a significant fraction of a day on a 2~GHz Pentium machine) due to the multi-dimensional numerical integrations involved, but need only be carried out once for a particular energy range. The cross-sections and rates are stored in a \texttt{ZEBRA/FZ} file for use during event generation; the interaction rates can be recalculated rapidly for a different target size or different flux with identical energy limits by reusing the stored cross-sections.  Various classes of reactions, or specific neutrino flavors, may be selectively processed or ignored via flags given on the command line.

Multi-dimensional integration of the rate and cross-section for each channel is performed by the Monte Carlo routine \texttt{MISER} or by cascaded Romberg integration.\cite{Recipes}  In both cases, a user-selectable tolerance on the estimated fractional error of the integral (0.3\% by default) is checked to determine whether convergence has been achieved. Some kinematic integration variables are transformed internally to accelerate convergence. The integral over neutrino energy necessary to compute total interaction rates can be done linearly or logarithmically.  To ensure consistency between calculated rates and simulated data, the same code is used for both cross-section calculations and event generation.

\subsection{Event Generation}

Once cross-sections and rates are calculated and stored, event generation proceeds very rapidly; when generated event vectors are written to disk, the program is I/O-limited. The user indicates the exposure time in units of seconds or years (for atmospheric and supernova fluxes, respectively) or units of luminosity (for neutrino beams).  A fixed number of events to generate may alternatively be specified. As in rate calculation, selected classes of reactions or neutrino flavors may be specified or suppressed.

Each reaction channel is tracked separately to determine how much simulation time elapses between events.  After determining which channel produces the next event, a neutrino energy is generated from the cumulative differential rate distribution for the reaction stored on the \texttt{FZ} file.  A set of transformed kinematic variables is created randomly within the allowed limits and the differential cross-section is calculated. The trial configuration is accepted or rejected by comparing a random number to the ratio of the differential cross-section to the maximum differential cross-section for the same neutrino energy encountered during the rate integration phase (a table of $\sigma_{max} \ \mathrm{vs.} \ E_{\nu}$ for each channel is also stored in the \texttt{FZ} file). If a particular set of kinematic variables is rejected, a new set is generated for the same reaction and neutrino energy until an accepted configuration is obtained. Using an accepted set of kinematic variables, four-vectors for the reaction's primary outgoing particles are generated. These particles are stepped through the target nucleus and/or decayed, if appropriate, and the final outgoing event vectors are written to disk.

For obvious reasons, the program is a voracious consumer of random numbers. The \texttt{RANLUX} package\cite{CernlibRanlux} is used everywhere (including external libraries, which are modified to call it instead of their native random number generators). The user should provide a 32-bit integer seed in the steering cards or on the command line to ensure that different jobs produce unique results.

The program writes its results in either (or both) of two formats as requested in the steering cards or on the command line: a human-readable text file or an \texttt{HBOOK} N-tuple.  The text file lists not only the initial and final-state tracks, but also the primary outgoing particles prior to nuclear interactions and decays. The reaction channel, elapsed simulation time, neutrino flux at the generated angle/energy and various internal or informational kinematic variables (N-tuple only) are also recorded. Total charged- and neutral-current cross-sections for each neutrino species and target combination are written to the \texttt{HBOOK} file in a separate N-tuple.

\section{TOWARD A UNIVERSAL MODEL}

\subsection{Motivation}

As outlined in the preceding sections, accurately modeling neutrino interactions presents a unique challenge,
combining the complexity of the nuclear many-body problem with the formidable uncertainties and sparse experimental guidance typical of neutrino physics. At the same time, discovery of neutrino oscillation has opened a new and inviting window on physics beyond the Standard Model, which ambitious future long-baseline experiments now being planned aim to exploit.

To date, each new experiment has had to solve the problem of simulating neutrino interactions independently. The literature contains a patchwork of disconnected and sometimes incompatible models. Piecing them together involves an equal mixture of intuition, folklore, and guesswork; presently, everyone working on the problem must make their own guesses in a vacuum, largely by trial and error. Without a common point of reference, there is no easy way to test the validity of one's assumptions and no avenue for one group's progress or ideas to reach others.  In short, there is virtually unlimited room for consolidation and incremental improvement in the state of the art, but no mechanism for it to occur. One can only imagine the chaos which would prevail in $p \overline{p}$ or $e^+ e^-$ physics if the LUND parton shower and string fragmentation code were not universally available and each experiment were instead forced to recreate something similar on their own, yet this is precisely the present situation in neutrino physics.

Further, as interest in neutrino physics continues to grow, more people enter the field and more future experiments are explored, the need for a general model only increases. On the other hand, the investment of time required to create a reliable simulation is prohibitive for those interested only in exploring future possibilities rather than operating an approved experiment. Most ``proprietary" code developed by a particular experimental collaboration tends to be specific to their own running conditions, and many groups are reluctant to release internal software. 

\subsection{The NUINT Working Group}

The first NUINT meeting was a watershed in the history of neutrino physics. In addition to bringing together a diverse group from across the spectrum of nuclear and particle physics to offer their specialized insights and expertise, it also united for the first time a large subset of the world's neutrino simulation experts, many of whom have labored in isolation for years. For me, this opportunity to compare notes and exchange ideas was the high-point of an already-fascinating workshop.

Agreement on the need for more comprehensive and widely available software tools was unanimously expressed, along with a willingness to work together toward them in the future.  The first step will be to compare our results in a number of agreed ``benchmark" cases.  This work will highlight the most important areas of ambiguity and uncertainty, and hopefully motivate renewed theoretical attention on long-neglected but essential points.  A central web site (\texttt{http://nuint.ps.uci.edu}) linked to the homepages of each participant will allow the working group and the community as a whole to track the progress of this effort and provide feedback.

For the longer term, the efforts of many will be required to produce a carefully-tested and universal model of neutrino interactions.  In addition to purely technical considerations, theoretical guidance and new experimental data will be vital.  Still, with the success of NUINT'01 and the promise of renewed and expanded collaboration punctuated and reinforced by future NUINT workshops, it is not too optimistic to hope that within a relatively few years, members of the neutrino physics community will finally have at their disposal a software tool equal to the challenges and opportunities they face.

\section{Acknowledgements}

I would like to thank T.~Barszczak, W.~Gajewski, A.~Habig, D.~Kielczewska, P.~Lipari, C.~McGrew, K.~Scholberg, C.~Walter, B.~Viren
and the IMB, Super-Kamiokande and K2K collaborations for important contributions and feedback during the on-going development
of the program.

%


\begin{thebibliography}{99}
\bibitem{IMB1} T.~J.~Haines et al., Phys.\ Rev.\ Lett.\ 57 (1986) 1986.
\bibitem{HainesThesis} T.~J.~Haines, University of California, Irvine Ph.D.\ Thesis (1986).
\bibitem{KamMc} M.~Nakahata et al., J.\ Phys.\ Soc.\ Jpn.\ 55 (1986) 3786.
\bibitem{DwcThesis}  D.~Casper, University of Michigan Ph.D.\ Thesis (1990).
\bibitem{IMB3} D.~Casper et al., Phys.\ Rev.\ Lett.\ 66 (1991) 2561; \\ 
                 R.~Becker-Szendy et al., Phys.\ Rev.\ D 46 (1992) 3720.
\bibitem{Pythia6} T.~Sj\"ostrand, P.~Ed\'en, C.~Friberg, L.~L\"onnblad, G.~Miu, S.~Mrenna and E.~Norrbin, 
Computer Phys.\ Commun.\ 135 (2001) 238. \\ (LU TP 00-30, hep-ph/0010017)
\bibitem{PDG} D.~E.~Groom et al., The European Physics Journal C15 (2000) 1.


\bibitem{Moniz:1971mt}
E.~J.~Moniz, I.~Sick, R.~R.~Whitney, J.~R.~Ficenec, R.~D.~Kephart and W.~P.~Trower,
Phys.\ Rev.\ Lett.\  26 (1971) 445.


\bibitem{Greiner} W.~Greiner and B.~M\"uller, {\it Gauge Theory of Weak Interactions} (Springer, Berlin 2000).


\bibitem{Smith:xh}
R.~A.~Smith and E.~J.~Moniz,
Nucl.\ Phys.\ B 43 (1972) 605. 
{[Erratum-ibid.\ B 101 (1975) 547]}.

\bibitem{Liu:1994dr}
K.~F.~Liu, S.~J.~Dong, T.~Draper and W.~Wilcox,
Phys.\ Rev.\ Lett.\ 74 (1995) 2172
[arXiv:hep-lat/9406007].

\bibitem{Ahrens:xe}
L.~A.~Ahrens et al.,
Phys.\ Rev.\ D 35 (1987) 785.

\bibitem{Pais}
A.~Pais, Annals Phys.\ 63 (1971) 361.


\bibitem{ReinSehgal}
D.~Rein and L.~M.~Sehgal,
Annals Phys.\  133 (1981) 79; \\
D.~Rein,
Z.\ Phys.\ C 35 (1987) 43.

\bibitem{Feynman:wr}
R.~P.~Feynman, M.~Kislinger and F.~Ravndal,
Phys.\ Rev.\ D 3 (1971) 2706.

\bibitem{Kim:1996az}
H.~C.~Kim, S.~Schramm and C.~J.~Horowitz,
Phys.\ Rev.\ C 53 (1996) 3131
[arXiv:nucl-th/9602009].

\bibitem{Alvarez-Ruso:1999dy}
L.~Alvarez-Ruso, E.~Oset, S.~K.~Singh and M.~J.~Vicente-Vacas,
Nucl.\ Phys.\ A 663 (2000) 837
[arXiv:nucl-th/9907109]; \\
S.~K.~Singh, M.~J.~Vicente-Vacas and E.~Oset,
Phys.\ Lett.\ B 416 (1998) 23.

\bibitem{Marteau:ur}
J.~Marteau, J.~Delorme and M.~Ericson,
Nucl.\ Phys.\ A 663 (2000) 783.


\bibitem{Rein:1982pf}
D.~Rein and L.~M.~Sehgal,
Nucl.\ Phys.\ B 223 (1983) 29.

\bibitem{Kelkar:1996iv}
N.~G.~Kelkar, E.~Oset and P.~Fernandez de Cordoba,
Phys.\ Rev.\ C  55 (1997) 1964
[arXiv:nucl-th/9609005].


\bibitem{Rein:cd}
D.~Rein,
Nucl.\ Phys.\ B 278 (1986) 61.

\bibitem{vmd}
M.~K.~Gaillard, S.~A.~Jackson and D.~V.~Nanopoulos,
Nucl.\ Phys.\ B 102 (1976) 326
[Erratum-ibid.\ B 112 (1976) 545]; \\
M.~S.~Chen, F.~S.~Henyey and G.~L.~Kane,
Nucl.\ Phys.\ B 118 (1977) 345. \\
P.~Marage et al., 
Z.\ Phys.\ C 35 (1987) 275.



\bibitem{Roe}
B.~P.~Roe, University of Michigan preprint UM-HE-94-10 (1994) [NuMI-29].

\bibitem{Leader:hm}
E.~Leader and E.~Predazzi,
Cambridge Monogr.\ Part.\ Phys.\ Nucl.\ Phys.\ Cosmol.\  4 (1996) 1.

\bibitem{Plothow-Besch:1992qj}
H.~Plothow-Besch,
Int.\ J.\ Mod.\ Phys.\ A 10 (1995) 2901; \\
H.~Plothow-Besch,
Comput.\ Phys.\ Commun.\  75 (1993) 396.  

\bibitem{BEBC}
A.~J.~Buras and K.~J.~F.~Gaemers, Nucl. Phys. B132 (1978) 249; \\
K.~Varnell et al., 
Z. Phys. C36 (1987) 1.

\bibitem{Ingelman:1996mq}
G.~Ingelman, A.~Edin and J.~Rathsman,
Comput.\ Phys.\ Commun.\ 101 (1997) 108
[arXiv:hep-ph/9605286].

\bibitem{Koba:ng}
Z.~Koba, H.~B.~Nielsen and P.~Olesen,
Nucl.\ Phys.\ B 40 (1972) 317; \\
D.~Levy,
Nucl.\ Phys.\ B 59 (1973) 583; \\
F.~Hayot and G.~Sterman,
Phys.\ Lett.\ B 121 (1983) 419.

\bibitem{Schmitz}
N. Schmitz in {\it Neutrino physics and astrophysics (Neutrino '88)} proceedings, J.~Schneps, T.~Kafka, W.~A.~Mann, P.~Nath, eds. (International Conference on Neutrino Physics and Astrophysics, 13th, Boston, Mass., Jun 5-11, 1988), 1989. 

\bibitem{Bodek:2002vp}
A.~Bodek and U.~K.~Yang, these proceedings
[arXiv:hep-ex/0203009].


\bibitem{Manjeet}
M.~Mudan, University of London, University College Ph.D.\ Thesis (1989).

\bibitem{Sick}
A.~Sick et al., Nucl.\ Phys.\ A 150 (1970) 631.

\bibitem{CERN-HERA}
V.~Flaminio, W.~G.~Moorhead, D.~R.~Morrison and N.~Rivoire,
CERN-HERA-83-01; \\
V.~Flaminio, W.~G.~Moorhead, D.~R.~Morrison and N.~Rivoire,
CERN-HERA-83-02; \\
V.~Flaminio, W.~G.~Moorhead, D.~R.~Morrison and N.~Rivoire,
CERN-HERA-84-01.

\bibitem{Karlsruhe}
G.~H\"ohler in {\it Landolt-B\"ornstein, Vol.\ I/9b2}, (Springer, Berlin 1999-2001).

\bibitem{OllsenYodh}
M.~G.~Ollsen and G.~B.~Yodh, Phys.\ Rev.\ 145 (1966) 1309.

\bibitem{Kaons}
B.~R.~Martin and M.~K.~Pidcock, Nucl.\ Phys.\ B 126 (1977) 266; \\
M.~Alston-Garnjost, R.~W.~Kenney, D.~L.~Pollard, R.~R.~Ross, R.~D.~Tripp and H.~Nicholson,
Phys.\ Rev.\ D 17 (1978) 2226.

\bibitem{rhoOmega}
W.~Deinet, H.~Mueller, D.~Schmitt, H.~M.~Staudenmaier, S.~Buniatov and E.~Zavattini,
Nucl.\ Phys.\ B  11 (1969) 495; \\
J.~Keyne et al.,
Phys.\ Rev.\ D 14 (1976) 28; \\
J.~Feltesse et al.,
Nucl.\ Phys.\ B 93 (1975) 242; \\
R.~D.~Baker et al.,
Nucl.\ Phys.\ B 156 (1979) 93.

\bibitem{Adler:qu}
S.~L.~Adler, S.~Nussinov and E.~A.~Paschos,
Phys.\ Rev.\ D 9 (1974) 2125
[Erratum-ibid.\ D 10 (1974) 1669].

\bibitem{fzic}
H.~J.~Moehring and J.~Ranft,
Z.\ Phys.\ C 52 (1991) 643; \\
G.~Battistoni, C.~Forti, J.~Ranft and S.~Roesler,
Astropart.\ Phys.\  7 (1997) 49.
[arXiv:hep-ph/9606485].

\bibitem{Ejiri}
H.~Ejiri,
Phys.\ Rev.\ C 48 (1993) 1442.




\bibitem{Barger}
V.~D.~Barger, K.~Whisnant, S.~Pakvasa and R.~J.~Phillips,
Phys.\ Rev.\ D 22 (1980) 2718; \\
V.~D.~Barger, K.~Whisnant and R.~J.~Phillips,
Phys.\ Rev.\ Lett.\  45 (1980) 2084.


\bibitem{Tauola} S.~Jadach, Z.~Was and J.~H.~Kuehn, Comp.\ Phys.\ Commun.\ 64 (1991) 275; \\
M.~Jezabek, Z.~Was, S.~Jadach, J.~H.~Kuehn, Comp.\ Phys.\ Commun.\ 70 (1992) 69.

\bibitem{Albright:1974ts}
C.~H.~Albright and C.~Jarlskog,
Nucl.\ Phys.\ B 84 (1975) 467.

\bibitem{Lipari:ut}
P.~Lipari and T.~Stanev,
Phys.\ Rev.\ D 44 (1991) 3543.


\bibitem{CernlibZebra}
CERN Program Library Long Write-ups Q100/Q101. 


\bibitem{CMT}
C.~Arnault, in {\it Computing in high energy and nuclear physics (CHEP 2000) : Proceedings}, M.~Mazzucato and M.~Michelotto, eds., (Padua, Italy, February 7-11, 2000) 

\bibitem{Spergel:1987ch}
D.~N.~Spergel, T.~Piran, A.~Loeb, J.~Goodman and J.~N.~Bahcall,
Science 237 (1987) 1471.

\bibitem{Recipes}
W.~H.~Press, S.~A.~Teukolsky, W.~T.~Vetterling and B.~P.~Flannery, {\it Numerical Recipes in Fortran 77, 2nd ed.} (Cambridge University Press, Cambridge 1992).

\bibitem{CernlibRanlux}
M.~Luscher,
Comput.\ Phys.\ Commun.\  79 (1994) 100 
[arXiv:hep-lat/9309020]; \\
F.~James,
Comput.\ Phys.\ Commun.\  79 (1994) 111
[Erratum-ibid.\  97 (1996) 357].


\end{thebibliography}
\end{document}